\documentclass
[reprint,prl,a4paper,amsfonts,amssymb,superscriptaddress,onecolumn]{revtex4}
\usepackage{amssymb}
\usepackage{amsfonts}
\usepackage{amsmath}
\usepackage{graphicx}
\usepackage{array}
\usepackage{SIunits}
\usepackage{ulem}
\newcommand{\upcitep}[1]{\textsuperscript{\textsuperscript{\citep{#1}}}}
\setcitestyle{open={},close={}}
\allowdisplaybreaks[3]

\begin{document}

\title{TMR transition and highly sensitive pressure sensors based on magnetic tunnel junctions with black phosphorus barrier}
\author{Henan Fang*}
\author{Qian Li}
\author{Mingwen Xiao*}
\author{Yan Liu}

\begin{abstract}
\noindent \textbf{ABSTRACT:} Black phosphorus is a promising material to serve as the barrier of magnetic tunnel junctions (MTJs) due to the weak van der Waals interlayer interactions. In particular, the special band features of black phosphorus may bring intriguing physical characteristics. Here, we study theoretically the effect of band gap tunability of black phosphorus on the MTJs with black phosphorus barrier. It is found that, the tunneling magnetoresistance (TMR) may achieve a transition from finite value to infinity owing to the variation of the band gap of black phosphorus. Combining with the latest experimental results of the pressure-induced band gap tunability, we further investigate the pressure effect of TMR in the MTJs with black phosphorus barrier. The calculations show that the pressure sensitivity can be quite high under appropriate parameters. Physically, the high sensitivity originates from the TMR transition phenomenon. To take advantage of the high pressure sensitivity, we propose and design a detailed structure of highly sensitive pressure sensors based on MTJs with black phosphorus barrier, whose working mechanism is basically different from the convential pressure sensors. The present pressure sensors possess four advantages and benifits: (1) high sensitivity, (2) well anti-interference, (3) high spatial resolution, and (4) fast response speed. Our study may advance new research area for both the MTJs and pressure sensors.
\end{abstract}

\maketitle
\section{Introduction}
Black phosphorus, as one kind of two-dimensional semiconductor, has received tremendous attentions owing to its two-dimensionality and the availability of advanced characterization techniques \upcitep{rfb1}. For practical applications, black phosphorus has a great potential in field effect transistors, photodetectors, batteries, etc \upcitep{rfb1}. In the field of spintronics, the weak van der Waals interlayer interactions make it possible for black phosphorus to serve as the barrier of magnetic tunnel junctions (MTJs). Up to now, the research of MTJs with black phosphorus barrier is rarely addressed \upcitep{rfb2}. In particular, the bandgap tunability of black phosphorus, which may play a key role in tunneling magnetoresistance (TMR), has never been adequately considered.

The band gap tunability of black phosphorus mainly manifests as two aspects: thickness (layer) dependence and pressure dependence. On one hand, the band gap of black phosphorus is highly dependent on the thickness \upcitep{rfb1,rf11,rf12,rf13}. For the monolayer black phosphorus, the band gap is about $2\, \mathrm{eV}$ \upcitep{rfb1,rf11,rf13,rf14}, while for the bulk black phosphorus, the band gap is only about $0.3\, \mathrm{eV}$ \upcitep{rfb1,rf11,rf13,rf15}. On the other hand, quite recently, S. Huang et al. investigated experimentally the pressure effect on the electronic structure of black phosphorus \upcitep{rfb3}. It is found that, the relative changes of band gap versus pressure depend on the number of the layers of black phosphorus, and can be expressed as follows:
\begin{equation}
\bigtriangleup E^{N}_{g}(P)= aP-\frac{\gamma_{0}}{2}(\sqrt{1+\frac{P}{P_{\textrm{coh}}}}-1)\cos(\frac{1}{N+1}\pi)
\end{equation}
where $P$ is the magnitude of the pressure, $N$ is the number of the layers, $a$ is a changing rate, $\gamma_{0}$ is the difference of overlapping integrals for conduction band and valence band under $0\, \mathrm{GPa}$, and $P_{\textrm{coh}}$ is the cohesive pressure. In ref \cite{rfb3}, both the experiments and the derivation of eq 1 are irrelevant to the substrate. That is, eq 1 is the physical properties of the black phosphorus alone. In addition, it is worth noting that eq 1 is valid only for the case that the thickness of black phosphorus is not greater than 50 nm. This is because, for bulk black phosphorus, there is additional in-plane compresive strain besides the normal strain \upcitep{rfb3}. The above band gap tunability may bring novel characteristics for the MTJs with black phosphorus barrier, and more importantly, those characteristics possibly lead to new applications.

Here, we investigate theoretically the effect of band gap tunability on the MTJs with black phosphorus barrier. As can be seen in the following, the band gap tunability may result in TMR transition phenomenon. Furthermore, through utilizing the TMR transition, we propose and design a kind of highly sensitive pressure sensors, and its working mechanism is entirely different from the conventional pressure sensors. Next, in order to help the readers to understand the TMR transition phenomenon, we shall briefly illustrate the physical background.

\section{Theoretical background}
In conventional theories, e.g. Julli\`{e}re's model and Slonczewski's model \upcitep{rf1,rf2}, the MTJs with half-metallic electrodes will obtain infinite TMR near zero bias and low temperature. However, the experimental TMR in such MTJs was far away from infinity \upcitep{rf3,rf4}, which is in contradiction with the conventional theories. In other words, the conventional theories are inapplicable to experiments. Recently, a new spintronic theory is developed by us to handle the MTJs with single-crystal barriers \upcitep{rf5}. The theory is founded on traditional optical scattering theory \upcitep{rf6}. Within it, the barrier is treated as a diffraction grating that will result in strong coherence to the tunneling electrons. So far, the theory has successfully explained almost all the important effects in the MgO-based MTJs, including barrier thickness effect \upcitep{rf5}, temperature effect \upcitep{rf7}, bias effect \upcitep{rf8}, etc. In particular, the theory works well when it is applied to MTJs with half-metallic electrodes, and it can clarify clearly the physical mechanism for the contradiction between the conventional theories and experiments \upcitep{rf9}. Meanwhile, the theory shows that, for the MTJs with half-metallic electrodes, the TMR can be finite or infinite, which depends on the relation between the Fourier transformation of the atomic potential of the barrier $v(\mathbf{K}_{h})$, the chemical potential of the electrodes $\mu$, and the half of the exchange splitting of the electrodes $\Delta$. At zero bias voltage limit, if $v(\mathbf{K}_{h}) > \Delta -\mu$, the TMR will be finite, and vice versa. In physics, the parameter $v(\mathbf{K}_{h})$ should be approximately proportional to the band gap of the barrier \upcitep{rf5,rf6}. As such, if the band gap of the barrier is tunable within a wide range, it may occur that, in the tunable range of the band gap, there is a critical point corresponding to $v(\mathbf{K}_{h}) = \Delta -\mu$. Of course, on one side of the critical point, $v(\mathbf{K}_{h}) > \Delta -\mu$, and on the other side, $v(\mathbf{K}_{h}) < \Delta -\mu$. In this case, the TMR can achieve a transition from finite value to infinity just around the critical point even the material of the barrier remains unchanged.

\begin{figure}[ht]
\centering
\includegraphics[scale=0.5]{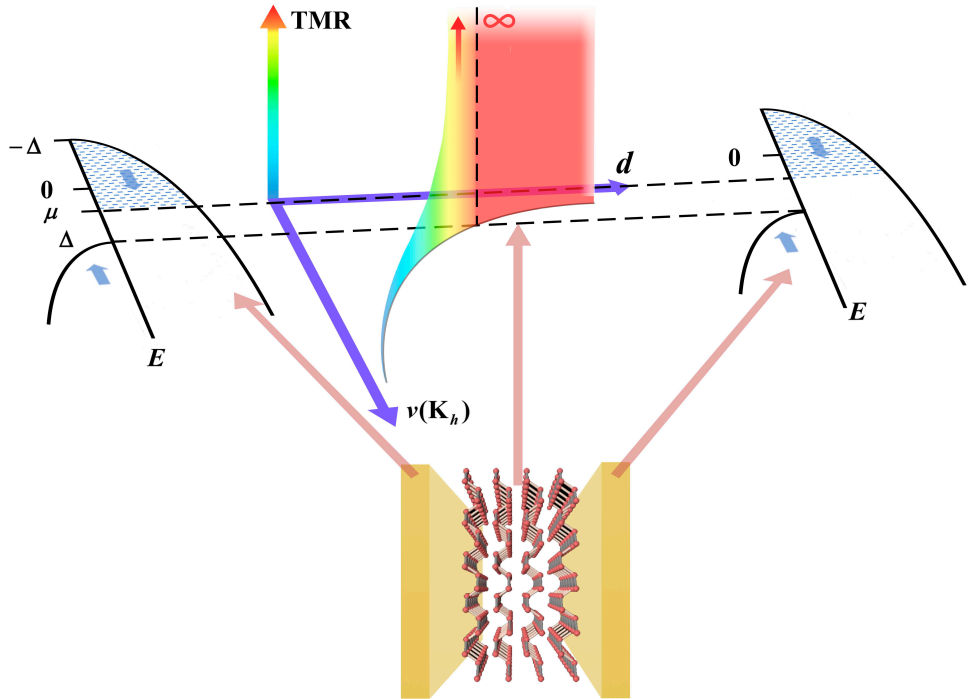}
\caption{The diagrammatic pictures for the thickness effect in the MTJs with black phosphorus barrier and half-metallic electrodes at zero bias limit. }
\end{figure}

\begin{figure}[ht]
\centering
\includegraphics[scale=0.45]{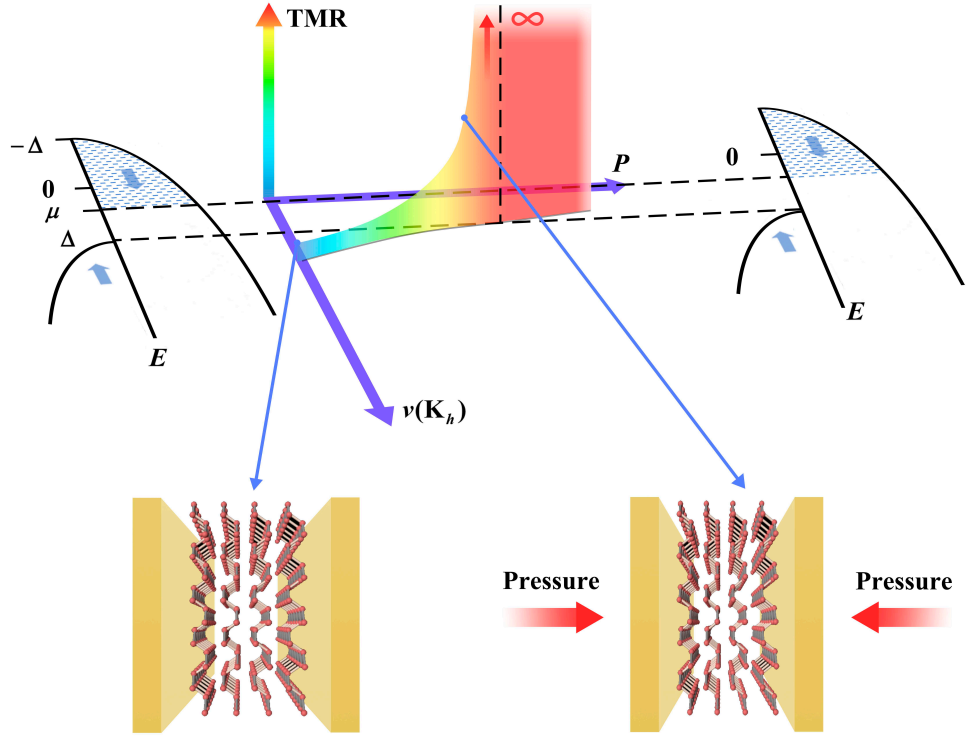}
\caption{The diagrammatic pictures for the pressure effect in the MTJs with black phosphorus barrier and half-metallic electrodes at zero bias limit. }
\end{figure}

As pointed out above, the band gap of black phosphorus is highly dependent on the thickness. The wide range of band gap modulation suggests that black phosphorus may be a good candidate as the barrier for realizing the scenario proposed above, i.e., the TMR can achieve a transition from finite value to infinity in the MTJs consisting of black phosphorus barrier with different thicknesses and half-metallic electrodes. The physical picture has been illustrated schematically in Figure 1 where the $z$-axes of spin for the two ferromagnetic electrodes are respectively chosen as their own. As shown in Figure 1, the TMR tends to infinity at a certain thickness which will be called as "critical thickness" in the following. This means that the TMR is extremely sensitive to the variation of barrier thickness when the barrier thickness is slightly less than the critical thickness. In physics, that is because the critical thickness just corresponds to the critical equation $v(\mathbf{K}_{h}) = \Delta -\mu$, as can been seen in Figure 1.

Let's now turn to the pressure effect. According to ref \cite{rfb3}, when $N\geq5$, the band gap of black phosphorus will decrease monotonically with pressure in the range of $0\, \mathrm{GPa}\sim1\, \mathrm{GPa}$, and the change is a few tens of $\mathrm{meV/GPa}$. Considering the case that, when the MTJs is without pressure, $v(\mathbf{K}_{h})$ is slightly larger than $\Delta -\mu$, and then a pressure is applied to the MTJs. In such case, $v(\mathbf{K}_{h})$ will decrease with the pressure and further tend to $\Delta -\mu$. Accordingly, the TMR will be extremely sensitive to the pressure, and the principle is essentially the same as the thickness effect. This provides a possible working mechanism for the highly sensitive pressure sensors based on MTJs with black phosphorus barrier, which has been illustrated schematically in Figure 2.

In the following, we shall calculate comprehensively the thickness effect and pressure effect of the MTJs with black phosphorus barrier (the theoretical model used for the calculations and the definition of TMR can be referred to the supporting information). According to the calculations, we design in detail the device prototype of the highly sensitive pressure sensors and discuss the advantages and benefits of it.

\section{Results and discussion}

The parameters in the calculations are set as follows. As stated in the theoretical background, the band gap $E_{g}$ of black phosphorus is highly dependent on the thickness. Until now, there is no unified expression of the dependence \upcitep{rf12,rf13}. Here, we adopt the expression of ref \cite{rf13}:
\begin{equation}
E_{g}= A\textrm{exp}(-BN)/N^{C}+D
\end{equation}
where $A = 1.71\, \mathrm{eV}$, $B = 0.17$, $C = 0.73$, $D = 0.40\, \mathrm{eV}$.
Since $v(\mathbf{K}_{h})$ should be approximately proportional to the band gap \upcitep{rf5,rf6}, we assume that the thickness dependence of $v(\mathbf{K}_{h})$ is the same as $E_{g}$. Considering the thickness of monolayer black phosphorus is $0.524\, \mathrm{nm}$ \upcitep{rf16}, we shall set
\begin{equation}
v(\mathbf{K}_{h})= 3.42\, \mathrm{eV}\textrm{exp}(-0.17d/0.524\, \mathrm{nm})/(d/0.524\, \mathrm{nm})^{0.73}+0.8\, \mathrm{eV}
\end{equation}
where $d$ is the thickness of the black phosphorus barrier. As for the pressure effect, according to ref \cite{rfb3}, the parameters in eq 1 are set as follows: $a = 0.18\, \mathrm{eV/GPa}$, $\gamma_{0} = 1.76\, \mathrm{eV}$, $P_{\textrm{coh}} = 1.4\, \mathrm{GPa}$. In addition, since the magnitude of the intralayer primitive translational vector $c = 4.3763\, \mathrm{{\AA}}$ \upcitep{rf16}, the magnitude of the intralayer reciprocal lattice vector $K_{h} = 2\pi/c = 1.436\times10^{10}\, \mathrm{m^{-1}}$.

\subsection{The thickness effect of TMR in MTJs with black phosphorus barrier}

\begin{figure}[ht]
\centering
\includegraphics[scale=0.5]{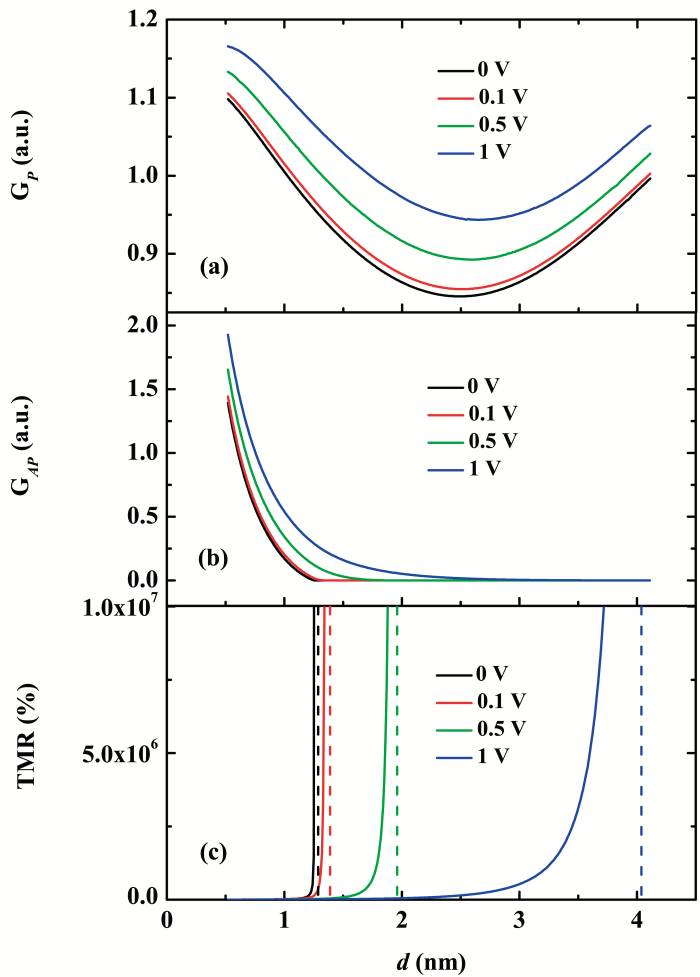}
\caption{(a)$G_{P}$, (b) $G_{AP}$ and (c) TMR as functions of barrier thicknesses under different bias voltage $V_{0} = 0\, \mathrm{V}, 0.1\, \mathrm{V}, 0.5\, \mathrm{V}$ and $1\, \mathrm{V}$. Here, the dashed lines denote the critical barrier thicknesses.}
\end{figure}

First, we would like to investigate the thickness dependence of conductances and TMR under different bias voltages. The theoretical results are depicted in Figure 3 where the chemical potential $\mu$ is $3\, \mathrm{eV}$, the half of the exchange splitting $\Delta$ is $5\, \mathrm{eV}$, and the bias $V_{0}$ is set sequentially as $0$, $0.1$, $0.5$ and $1\, \mathrm{V}$. As shown in Figure 3a, $G_{P}$ varies non-monotonically with the barrier thickness, which originates from the interference among the diffracted waves, as pointed out in ref \cite{rf5}. Unlike $G_{P}$, $G_{AP}$ firstly decreases with the barrier thickness and tends to zero at a certain thickness, as depicted in Figure 3b. It can be explained as follows: according to eq 3, $v(\mathbf{K}_{h})$ decreases monotonously with barrier thickness. Physically, the smaller the $v(\mathbf{K}_{h})$, the less energy the tunneling electrons will acquire. Therefore, the decreasing $v(\mathbf{K}_{h})$ will make less tunneling electrons get enough energy to transit from spin-up band into spin-down band. For the case of zero bias limit, as the barrier thickness increases, there will be a critical thickness ($d \approx 1.26\, \mathrm{nm}$) that corresponds to $v(\mathbf{K}_{h}) = \Delta -\mu$. At this critical thickness, $G_{AP}$ tends to zero because the energy $v(\mathbf{K}_{h})$ will be insufficient for the incident spin-up electrons to transit into spin-down band. For the finite bias case, the larger the bias voltage, the greater the critical thickness. This is because the critical thickness corresponds to $v(\mathbf{K}_{h}) = \Delta -\mu-V_{0}$ now, which has been illustrated schematically in Figure 4. From Figure 4, it can be seen that larger bias can reduce the energy required for the spin-up electrons to transit into spin-down band, i.e., smaller $v(\mathbf{K}_{h})$ is needed. According to eq 3, the critical thickness will be greater. Consequently, the TMR will achieve a transition from finite value to infinity around those critical thicknesses, as shown in Figure 3c. This is just the critical thickness phenomenon presented in Figure 1.

\begin{figure}[ht]
\centering
\includegraphics[scale=0.6]{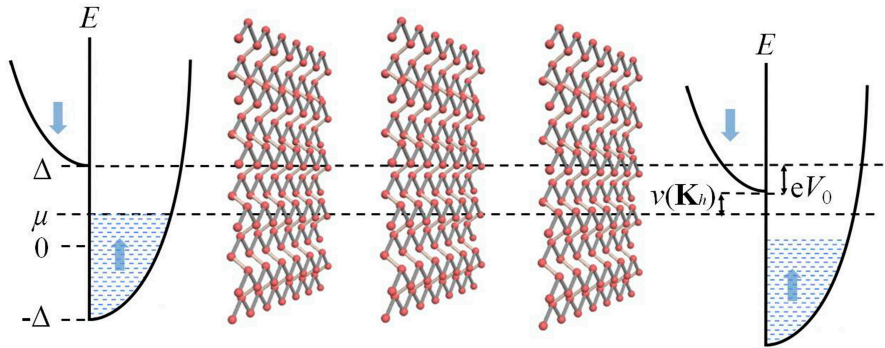}
\caption{An illustration of the potential of the MTJs at finite bias case. Here, the $v(\mathbf{K}_{h})$ corresponds to the critical equation $v(\mathbf{K}_{h}) = \Delta -\mu-V_{0}$.}
\end{figure}

\begin{figure}[ht]
\centering
\includegraphics[scale=0.5]{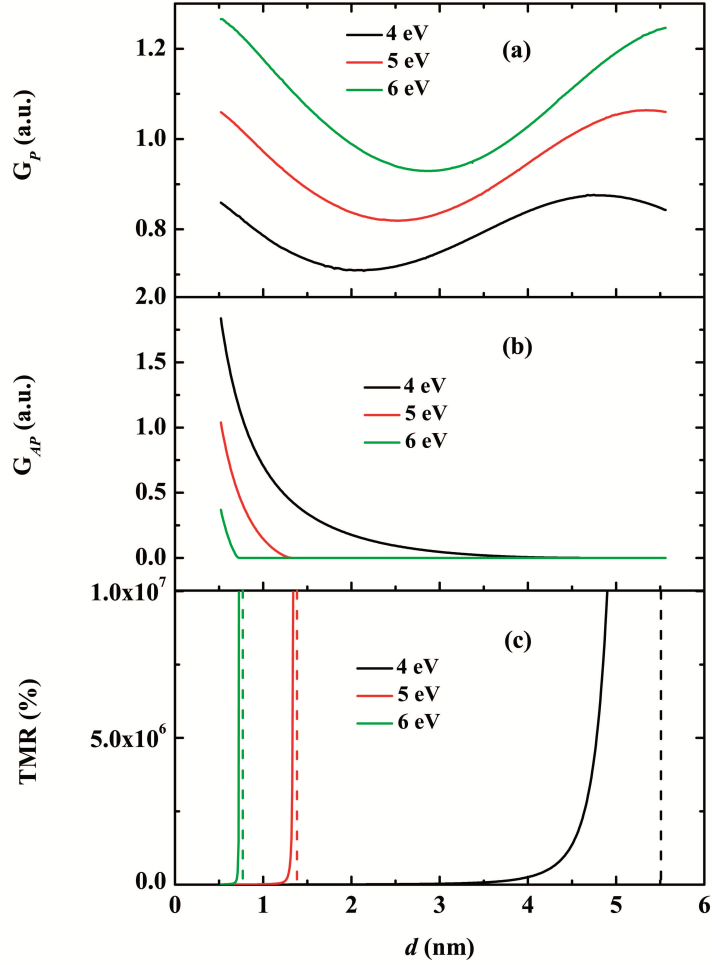}
\caption{(a)$G_{P}$, (b) $G_{AP}$ and (c) TMR as functions of barrier thicknesses under different half of the exchange splitting $\Delta = 4\, \mathrm{eV}, 5\, \mathrm{eV}$, and $6\, \mathrm{eV}$. Here, the dashed lines denote the critical barrier thicknesses.}
\end{figure}

Secondly, we shall study the thickness dependence of conductances and TMR under different half of the exchange splitting. The theoretical results are depicted in Figure 5 where the chemical potential $\mu$ is $3\, \mathrm{eV}$,  the bias is $0.1\, \mathrm{V}$, and the half of the exchange splitting $\Delta$ is set sequentially as $4$, $5$, and $6\, \mathrm{eV}$. As can be seen in Figure 5b,c, the critical thickness decreases with the half of the exchange splitting $\Delta$. This can be easily understood: according to the critical equation $v(\mathbf{K}_{h}) = \Delta -\mu-V_{0}$, when $\Delta$ increases, the $v(\mathbf{K}_{h})$ corresponding to the critical thickness increases as well. From eq 3, it can be known that the critical thickness will decrease with $\Delta$.

\begin{figure}[ht]
\centering
\includegraphics[scale=0.5]{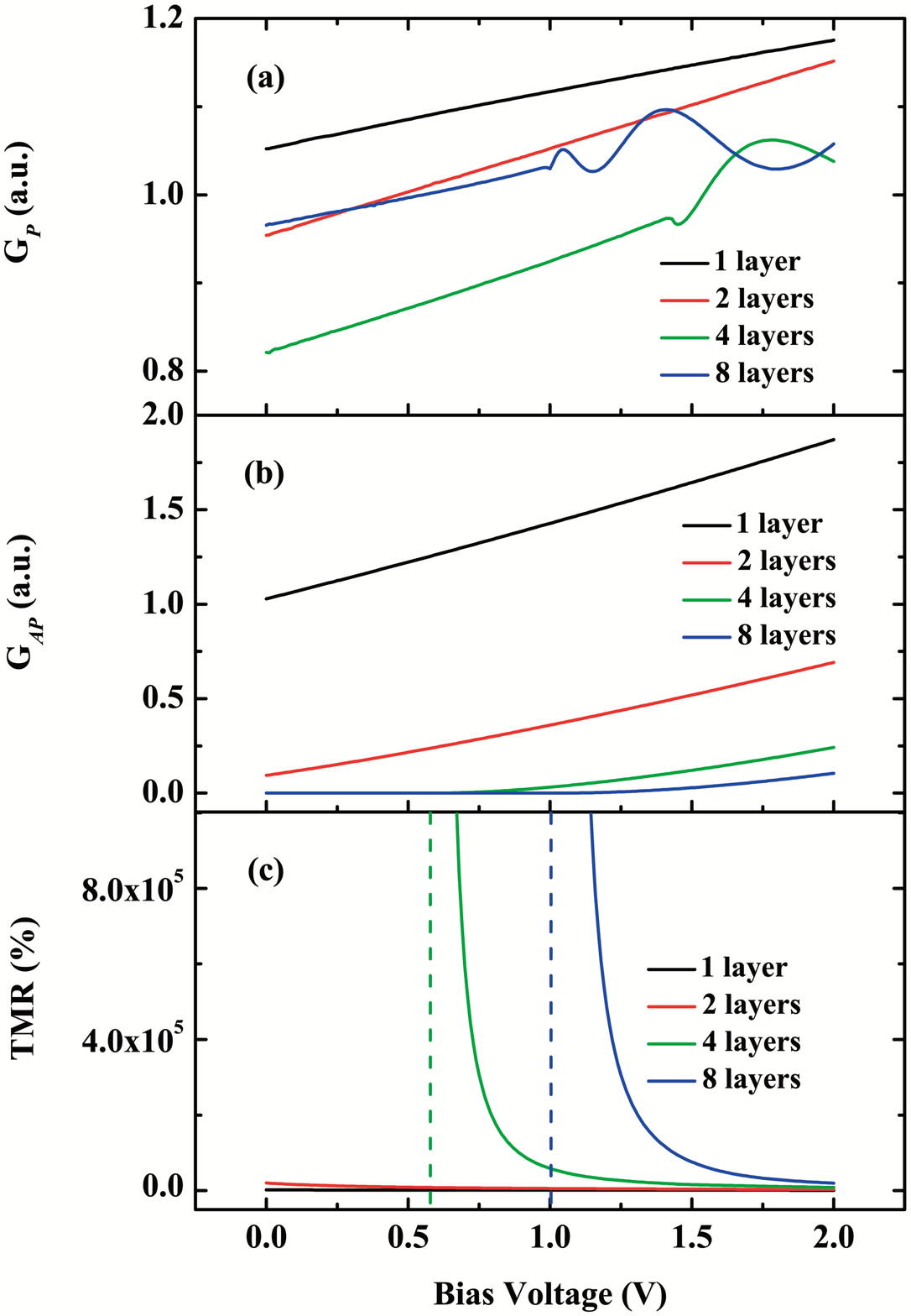}
\caption{(a)$G_{P}$, (b) $G_{AP}$ and (c) TMR as functions of bias voltage under different number of layers of black phosphorus barrier $N = 1, 2, 4$, and $8$. Here, the dashed lines denote the critical bias voltage.}
\end{figure}

From Figures 3 and 5, it suggests that the critical thicknesses can be regulated by both the bias voltage and the parameters of the half-metallic ferromagnetic electrodes, which wholly roots from the critical equation $v(\mathbf{K}_{h}) = \Delta -\mu-V_{0}$. In the critical thickness phenomenon, the right side of the critical equation remains unchanged, and the left side of the critical equation, i.e. $v(\mathbf{K}_{h})$, varies to cross the critical point. Conversely, if the left side of the critical equation remains unchanged, and the right side of the critical equation, i.e. the bias voltage $V_{0}$ or the half of the exchange splitting $\Delta$, varies to cross the critical point, there will also appear critical phenomenon. In practical applications, the bias voltage is much easier altered than the parameters of the half-metallic ferromagnetic electrodes. Videlicet, the critical bias phenomenon can be easier to be observed. Thereupon, we will discuss the bias dependence of conductances and TMR under different number of layers of black phosphorus barrier. The theoretical results are depicted in Figure 6 where the chemical potential $\mu$ is $3\, \mathrm{eV}$,  the half of the exchange splitting $\Delta$ is $5\, \mathrm{eV}$, and the number of the layers of black phosphorus barrier $N$ is set sequentially as $1$, $2$, $4$ and $8$. As shown in Figure 6a, when $N = 4$ and $N = 8$, there are critical points located around $1.43\, \mathrm{V}$ and $1\, \mathrm{V}$ respectively for $G_{P}$. On the right side of the critical points, $G_{P}$ oscillates with the bias voltage, whereas on the left side $G_{P}$ dose not. It can be explained as follows: in the present case, the oscillation of $G_{P}$ originates from the oscillation term of $\cos [({p}_{+}^{z}-{p}_{-}^{z})d]$ that belongs to the channel of $T_{\uparrow \uparrow }$. It can be deduced from eq 3 of the supporting information that, on the right side of the critical points, ${p}_{-}^{z}$ will be real in all integral regions for the channel of $T_{\uparrow \uparrow }$; on the left side of the critical points, ${p}_{-}^{z}$ will be imaginary in some integral regions, which will damage the oscillation property of $\cos [({p}_{+}^{z}-{p}_{-}^{z})d]$. More importantly, it can be seen from Figure 6c that, for $N = 4$ and $N = 8$, TMR achieves a transition from finite value to infinity around $V_{0} = 0.57\, \mathrm{V}$ and $V_{0} = 1\, \mathrm{V}$ respectively. The above two values of $V_{0}$ meet the critical equation $v(\mathbf{K}_{h}) = \Delta -\mu-V_{0}$, and this is just the critical bias phenomenon mentioned above. For the cases of $N = 1$ and $N = 2$, because $\Delta -\mu-V_{0}$ is always smaller than $v(\mathbf{K}_{h})$, so the critical bias phenomenon will not happen. In a word, the critical phenomenon only can happen when critical equation $v(\mathbf{K}_{h}) = \Delta -\mu-V_{0}$ is met in the variable range.

\subsection{The pressure effect of TMR in MTJs with black phosphorus barrier}

\begin{figure}[ht]
\centering
\includegraphics[scale=0.5]{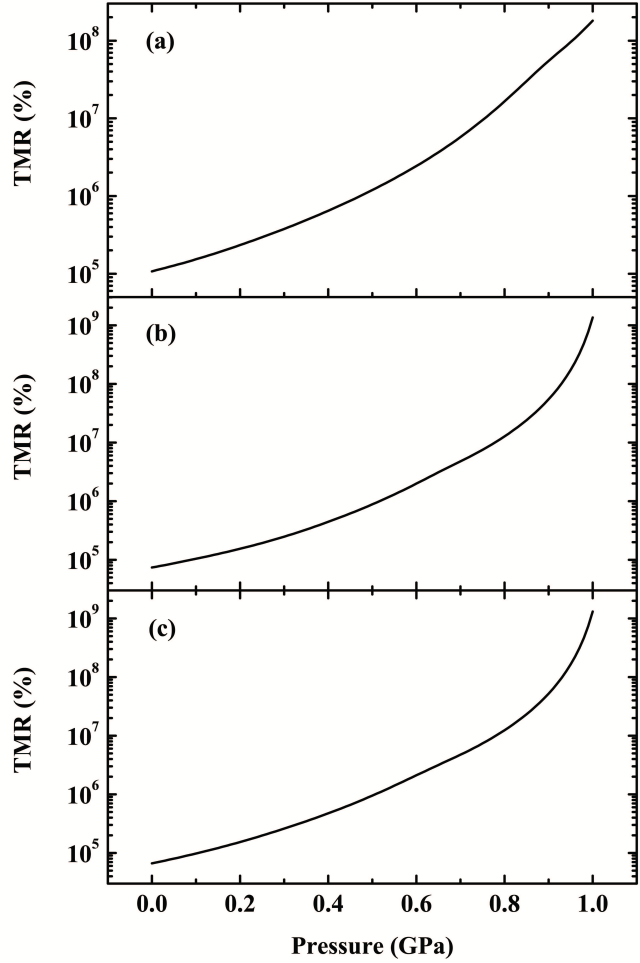}
\caption{TMR as functions of pressure under different number of layers of black phosphorus barrier (a)$N = 5$, (b)$N = 10$ and (c)$N = 20$. }
\end{figure}

As examples, we calculate respectively the pressure effect of TMR for three different thicknesses of black phosphorus barrier, i.e. $N = 5$, $N = 10$ and $N = 20$. The parameters are chosen to satisfy the condition that, when the MTJs is without pressure, $v(\mathbf{K}_{h})$ is slightly larger than $\Delta -\mu-V_{0}$. For the case of $N = 5$, $\mu = 3\, \mathrm{eV}$, $\Delta = 4.2\, \mathrm{eV}$ and $V_{0} = 70\, \mathrm{meV}$; for the case of $N = 10$, $\mu = 3\, \mathrm{eV}$, $\Delta = 3.8\, \mathrm{eV}$ and $V_{0} = 50\, \mathrm{meV}$; for the case of $N = 20$, $\mu = 3\, \mathrm{eV}$, $\Delta = 3.7\, \mathrm{eV}$ and $V_{0} = 70\, \mathrm{meV}$. The theoretical results are depicted in Figure 7 where the vertical is exponential type. As shown in Figure 7, for all of the three cases, the TMR increases with the pressure even more rapidly than exponential form. This is because, according to eq 1, the band gap of black phosphorus of these three thicknesses all decrease monotonically with pressure within the range $0\sim1\, \mathrm{GPa}$. In this sense, the monotonically decreasing range of the pressure-bandgap curve should be the theoretical range of detection. From eq 1, it can be derived that, the greater the number of layers, the larger the theoretical range of detection: for the number of layers $N = 5$, the theoretical range of detection is about $0\sim1.8\, \mathrm{GPa}$; for $N = 10$, the theoretical range of detection is about $0\sim2.5\, \mathrm{GPa}$; for $N = 20$, the theoretical range of detection is about $0\sim2.8\, \mathrm{GPa}$. It is worth noting that, the range of the experimental data in ref \cite{rfb3} only includes $0\sim2.5\, \mathrm{GPa}$, and therefore, the theoretical range of detection for $N = 20$ is only theoretically derived from eq 1. In the present case, the sensitivity can be defined as $(\mit\Delta \textrm{TMR}/\textrm{TMR}_{0})/\mit\Delta P$, where $\mit\Delta \textrm{TMR}$ is the change in TMR, $\textrm{TMR}_{0}$ is the TMR under no pressure, and $\mit\Delta P$ is the change in pressure. It can be deduced from Figure 7 that the highest sensitivity can reach $8.53\times10^{2}\, \mathrm{MPa}^{-1}$ at $1\, \mathrm{GPa}$ for the case of $N = 20$. Such high sensitivity originates from the TMR transition at the critical point, just as displayed in Figure 2. In principle, the present pressure sensors can reach arbitrary high sensitivity if the measuring range can be sacrificed. This is because, at the critical point of the TMR transition, TMR has a infinite derivative as well as sensitivity. In theory, we can let $\Delta -\mu-V_{0}$ tend to $v(\mathbf{K}_{h})$ infinitely, which means that the sensitivity can be arbitrarily high. The above results declare that, the MTJs with black phosphorus barrier indeed can be a potential system as highly sensitive pressure sensors.

\subsection{The design of highly sensitive pressure sensors}
Pressure sensors have a wide range of applications, such as transportation, automobile industries and medical field \upcitep{rfb4}. They constitute an important component of the field of microelectromechanical systems (MEMS), and have a highest market share among all the kinds of MEMS sensors \upcitep{rfb5}. The performance of pressure sensors can be assessed from many aspects, nevertheless, it is no doubt that the sensitivity is a vital parameter \upcitep{rfb6}. In particular, pressure sensors with high sensitivity are urgently needed in nanotechnology. When the sensitivity is high, well anti-interference, high spatial resolution and fast response speed are usually necessary. Therefore, the requirements of high sensitivity, well anti-interference, high spatial resolution and fast response speed are put forward for pressure sensors. Unfortunately, the conventional types of pressure sensors (e.g. piezoresistive-type sensors and capacitance-type sensors) can not completely meet the above requirements \upcitep{rfb7}. This suggests that novel sensing technology should be proposed to improve the corresponding performances.

On the other hand, the applications of MTJs are mainly focused on magnetic memory storage cells and magnetic field sensors. Besides the above familiar applications, M. L\"{o}hndorfa et al. develop a type of pressure sensors based on MTJs \upcitep{rfb7,rfb8}. Its working mechanism utilizes the inverse magnetostrictive effect (Villary effect), i.e., the shear pressure leads to a change in the magnetization direction of the free layer. In addition, the pressure sensitivity is described by the gauge factor: $GF = (\Delta R/R)/\Delta\varepsilon$, where $\Delta R/R$ is the relative change in resistance and $\Delta\varepsilon$ is the relative change in strain. Subsequently, a certain amount of research efforts have been devoted to improve the performances, especially the $GF$, of such pressure sensors \upcitep{rfb9,rfb10,rfb11,rfb12}. These works initiate a new research direction for both the MTJs and pressure sensors, and indicate that the pressure sensors based on MTJs have great potential to satisfy the requirements mentioned above. However, owing to the working mechanism, this type of pressure sensors can only measure the shear pressure, and is not applicable to the measurement of normal pressure. That confines the applications in certain fields.

\begin{figure}[ht]
\centering
\includegraphics[scale=0.35]{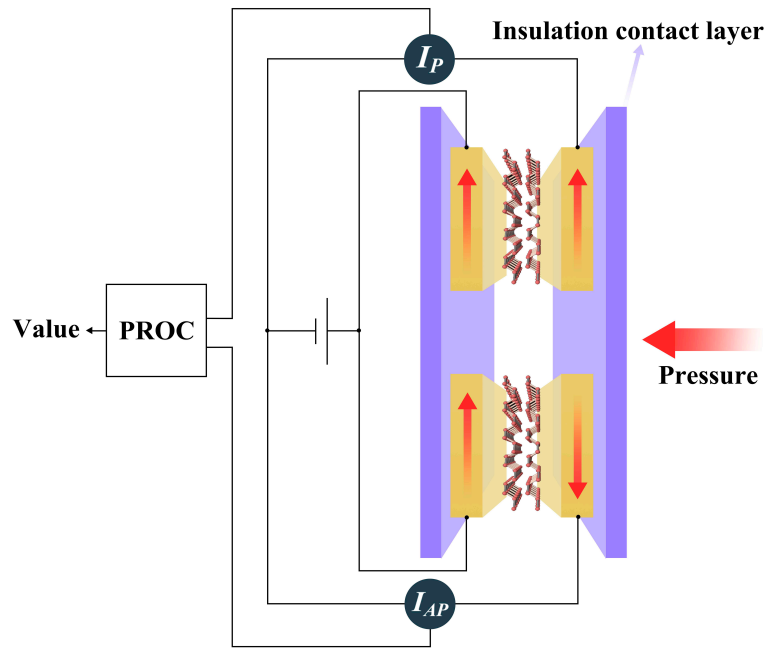}
\caption{Design diagram of the highly sensitive pressure sensors based on MTJs with black phosphorus barrier.}
\end{figure}

By this token, there needs a new kind of MTJs-based pressure sensors which can measure the normal pressure. The present calculations of the pressure effect demonstrate that the MTJ with black phosphorus barrier is good candidate to satisfy the need. Different from the previous MTJs-based pressure sensors, the working mechanism now is the TMR transition due to the pressure effect of black phosphorus. According to ref \cite{rfb3}, the pressure-induced variation of band gap roots from the decrease of the interlayer distance. In other words, the pressure in the normal direction leads to the variation of TMR. Therefore, the pressure sensors based on MTJs with black phosphorus barrier is suitable for measuring the normal pressure. In order to facilitate the practical application, here, a detailed structure of the device is designed, which is displayed in Figure 8. It can be seen from Figure 8 that, two MTJ with different magnetization configurations are in parallel connection to output parallel current and antiparallel current respectively. Such design has two advantages: (1) the parallel current $I_{P}$ and antiparallel current $I_{AP}$ can be simultaneously obtained, which can improve the response speed. (2) If there exists noise, it can only influence the magnitudes of $I_{P}$ and $I_{AP}$, but has no influence on the magnitude of TMR. In other words, the present pressure sensors has well anti-interference.  In addition, nanometer scale MTJs can lead to miniaturized pressure sensors with high spatial resolution \upcitep{rfb7}. Including the high sensitivity mentioned above, the present pressure sensors possess totally four advantages and benefits: (1) high sensitivity, (2) well anti-interference, (3) high spatial resolution, and (4) fast response speed. Besides, the completely new working mechanism may bring novel research area for both the pressure sensors and MTJs.

\section{Conclusions}
In this paper, we have studied the thickness effect and pressure effect of the MTJs with black phosphorus barrier. It is found that there exists TMR transition phenomenon: the TMR will achieve a transition from finite value to infinity around the critical thickness. The critical thickness can be regulated by both the bias voltage and half of the exchange splitting of electrodes. For the range of number of the black phosphorus layers $5\leq N \leq20$, the TMR increases with the pressure more rapidly than exponential form under appropriate parameters, and the sensitivity can be as high as $8.53\times10^{2}\, \mathrm{MPa}^{-1}$ at $1\, \mathrm{GPa}$ for the case of $N = 20$. Such high sensitivity originates from the TMR transition at the critical point. Furthermore, according to the calculations of the pressure effect, we design in detail the structure of the highly sensitive pressure sensors, and the design diagram is given. Finally, the advantages and benefits of the present pressure sensors are discussed.

By the way, the present work should not be confined within the MTJs with black phosphorus barrier. It can be extended to the MTJs with other band-tunable barrier, such as $\textrm{MoS}_{2}$ \upcitep{rfb13,rfb14}.

\section{S\lowercase{upporting information}}
The theoretical model for magnetic tunnel junctions with single crystal barrier

\section{N\lowercase{otes}}
The authors declare no competing interests.

\section{Acknowledgments}
This work is supported by the National Natural Science Foundation of China (11704197), the Natural Science Foundation of Nanjing University of Posts and Telecommunications (NY220184, NY221066).

\section{Author information}
\noindent \textbf{Corresponding authors}\\
\textbf{Henan Fang} - \textit{College of Electronic and Optical Engineering, Nanjing University of Posts and Telecommunications, Nanjing 210023, China};\\
Email: fanghn@njupt.edu.cn\\
\textbf{Mingwen Xiao} - \textit{Department of Physics, Nanjing University, Nanjing 210093, China};\\
Email: xmw@nju.edu.cn\\

\noindent \textbf{Authors}\\
\textbf{Qian Li} - \textit{College of Electronic and Optical Engineering, Nanjing University of Posts and Telecommunications, Nanjing 210023, China}\\
Email: lq842362507@163.com\\
\textbf{Yan Liu} - \textit{College of Electronic and Optical Engineering, Nanjing University of Posts and Telecommunications, Nanjing 210023, China}\\
Email: liuy9698@163.com

\clearpage
TOC graphic:
\begin{figure}[ht]
\centering
\includegraphics[scale=0.5]{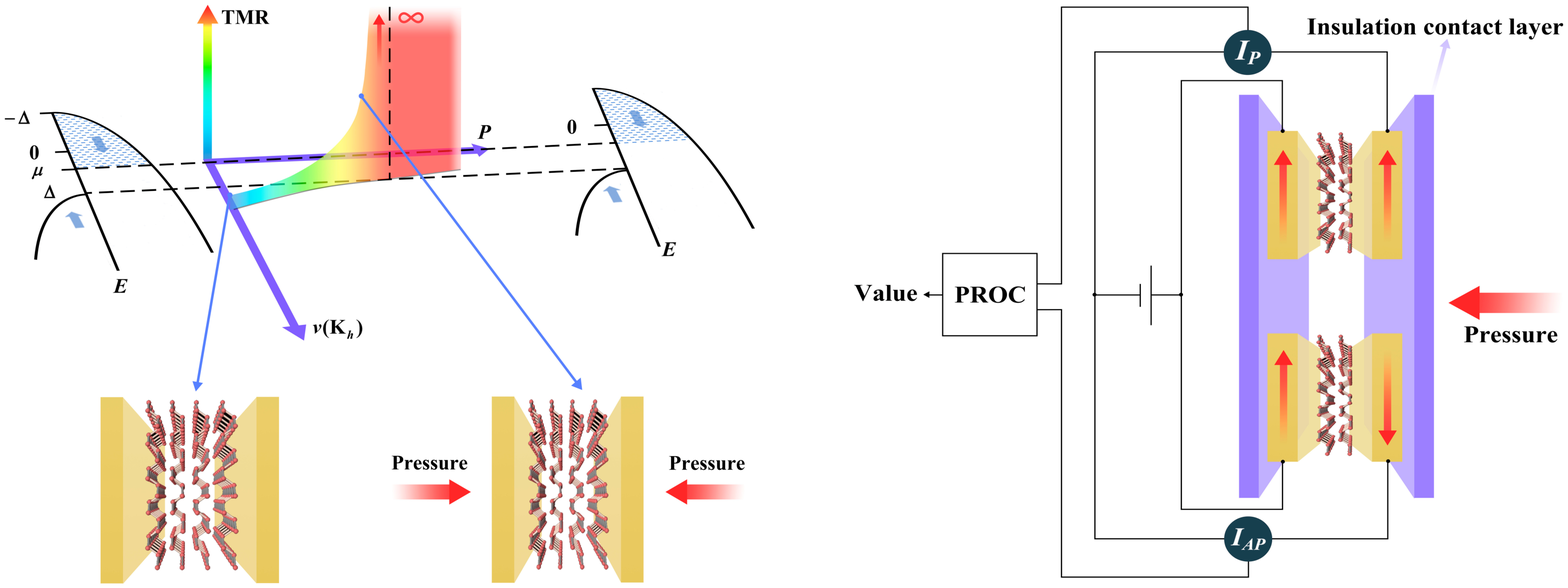}
\end{figure}

\end{document}